\documentclass[letterpaper,floatfix,aps,prb,amsmath,amssymb,twocolumn,
showpacs]{revtex4-1}

\usepackage{epsfig}
\usepackage{verbatim}
\usepackage{epsf}
\usepackage{array}
\usepackage[pdfauthor={Gianluigi Catelani},
            pdftitle={Parity switching and decoherence by quasiparticles in single-junction transmons}]{hyperref}

\hypersetup{pdfstartview={XYZ null null 1.09}}

\newcommand{\be}{\begin{equation}}
\newcommand{\ee}{\end{equation}}
\newcommand{\bea}{\begin{eqnarray}}
\newcommand{\eea}{\end{eqnarray}}

\renewcommand{\Re}{\mathrm{Re}\,}
\renewcommand{\Im}{\mathrm{Im}\,}

\newcommand{\Tr}{\mathrm{Tr}}
\newcommand{\eref}[1]{Eq.~(\ref{#1})}
\newcommand{\esref}[1]{Eqs.~(\ref{#1})}
\newcommand{\rref}[1]{(\ref{#1})}

\newcommand{\ocite}[1]{Ref.~\onlinecite{#1}}

\newcommand{\qp}{\mathrm{qp}}

\newcommand{\EZO}{\omega_{10}}

\begin{document}

\title{Parity switching and decoherence by quasiparticles in single-junction transmons}

\author{G. Catelani}

\affiliation{Peter Gr\"unberg Institut (PGI-2), Forschungszentrum J\"ulich, 52425 J\"ulich, Germany}

\begin{abstract}
The transmon superconducting qubit is being intensely investigated as a promising approach for the physical implementation of quantum information processing, and high quality factors of order $10^6$ have been achieved both in two- and three-dimensional architectures. These high quality factors enable detailed investigations of decoherence mechanisms. An intrinsic decoherence process originates from the coupling between the qubit degree of freedom and the quasiparticles that tunnel across Josephson junctions. In a transmon, tunneling of a single quasiparticle is associated with a change in parity. Here we present the theory of the parity-switching rates in single-junction transmons and compare it with recent measurements. We also show that parity switching can have an important role in limiting the coherence time.
\end{abstract}

\date{\today}


\maketitle

\section{Introduction}

State-of-the-art superconducting qubits have recently reached coherence times four orders of magnitude longer than those obtained
in pioneering experiments with Cooper pair boxes,\cite{cpb} and are close to meeting (or may have already met) the requirements for quantum error correction to be implemented.\cite{divincenzo,science}
Part of this significant improvement can be attributed to the development
of new qubit designs; the transmon,\cite{transmon} together with its so-called 3D implementation,\cite{paik} is at present one
of the most promising designs for quantum information applications. The long coherence times achieved, moreover, make possible
to study with increasing precision the roles of different decoherence processes, such as quasiparticle effects\cite{prl} and photon shot noise dephasing.\cite{psn}
In this paper, we consider in detail the quasiparticle mechanism of decoherence in a single-junction transmon.

\begin{figure}[b]
 \includegraphics[width=0.48\textwidth]{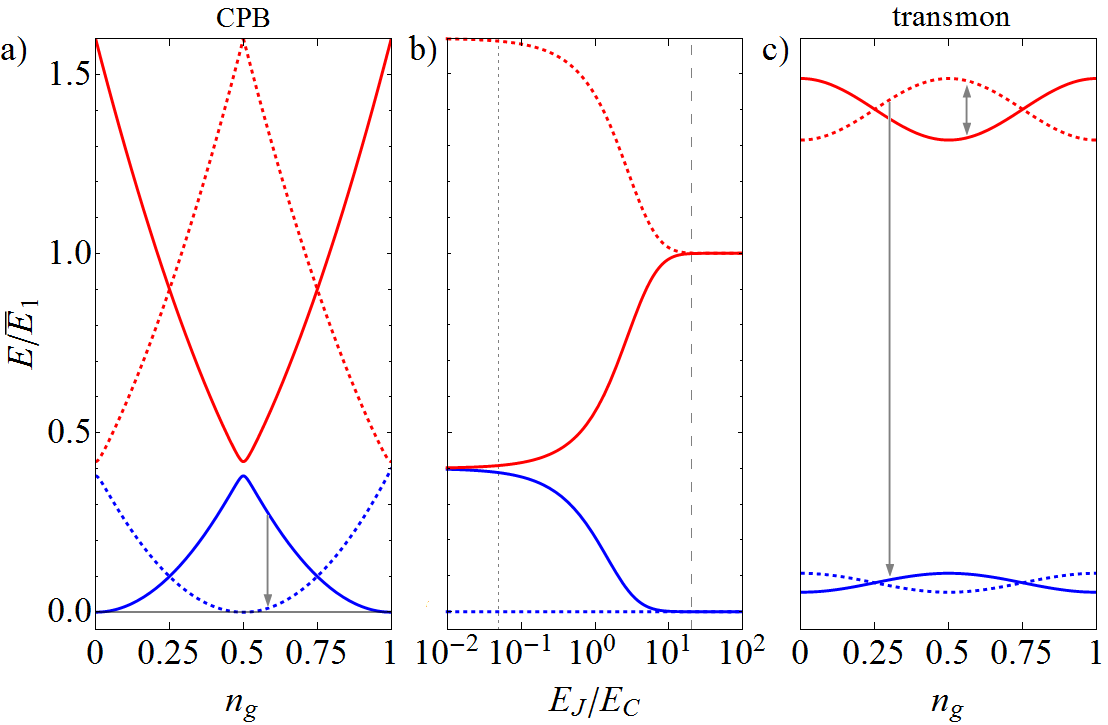}
 \caption{(color online): Solid (dotted) lines are used for even (odd) parity states in all panels. Arrows denote possible quasiparticle induced transitions. a) energy levels as functions of $n_g$ for a Cooper pair box with $E_J/E_C=0.05$. Energy is normalized by the average energy of the third and fourth state, $\bar{E}_1 = (E_1^e+E_1^o)/2$, at $n_g=1/2$. b) energy of the four lowest states at $n_g=1/2$ as function of the ratio $E_J/E_C$. The vertical scale is the same as in panel a).
 The vertical dotted line is at $E_J/E_C=0.05$, while the dashed line at $E_J/E_C=20$ demarcates the transmon regime to its right. c) schematic representation (energies not to scale) of the energy levels as functions of $n_g$ for a transmon. }
 \label{figlev}
\end{figure}

The transmon was originally introduced to decrease the sensitivity to charge noise of the Cooper pair box (CPB). In the latter, the largest energy scale is the charging energy $E_C$, which leads to the dominant parabolic dependence of the energy levels on (dimensionless) gate voltage $n_g$, see left panel in Fig.~\ref{figlev}. The qubits states are superpositions of states with the same parity -- that is, states which differ by tunneling of a Cooper pair; such a pair-tunneling process does not change the parity (even or odd) of the number of electrons which have tunneled through the junction. The sensitivity to charge noise manifests itself in Fig.~\ref{figlev}a) as a large variation in the energy of the levels for a small change in $n_g$. That is why the qubit must be operated at the optimal point of minimum energy difference (given by the Josephson energy $E_J$). By increasing $E_J$ the separation between same-parity levels increase, while they approach in energy the nearby levels with opposite parity, see Fig.~\ref{figlev}b). At the same time, the dependence of energy on $n_g$, and thus the sensitivity to charge noise, weakens [Fig.~\ref{figlev}c)]. The same diagram help us understanding why the transmon is also less disturbed by
so-called ``quasiparticle poisoning'':\cite{lutchyn} in the CPB, tunneling of a single excitation through the junction changes the parity of the state, bringing the system outside the
qubit subspace [see arrow in Fig.~\ref{figlev}a)]. In contrast, each transmon logical qubit state consists of two states: the two lowest energy states of opposite parity correspond to one qubit state, and the two states at higher energy to the other qubit state. Quasiparticle tunneling events always change the parity, but not necessarily the qubit state if they cause transitions between physical states corresponding the the same logical state [see, \textit{e.g.}, the short arrow in Fig.~\ref{figlev}c)]; we call these transitions parity-switching events. Those events in which the energy change is large lead to relaxation of the qubit [long arrow in Fig.~\ref{figlev}c)].

The relaxation of superconducting qubits induced by quasiparticles has been the considered in a number of recent theoretical and experimental works.\cite{martinis,prl,prb,marth2,paik,lenander,corcoles} Bounds on the parity switching rates
were placed in Refs.~\onlinecite{psb1,sun}, while direct measurements of those rates have been performed in \ocite{riste}.
For the theoretical description of the qubit, the multi-level physical system is in general reduced\cite{noteswt} to a two-level system. However, for the transmon this reduction does not provide a sufficiently detailed description; it misses, for example, the parity-switching events described above. Here we explicitly keep the four lowest levels: this enables us to study the parity-switching rates, compare the theoretical results with recent measurements, and elucidate the role of parity switching in the transmon dephasing.

The paper is organized as follows: in the next Section we introduce the effective Hamiltonian
of the single-junction transmon, including its interaction with quasiparticles. In Sec.~\ref{sec:re} we consider
phenomenological rate equations that can be used to describe relaxation and parity switching; microscopic expressions for the rates are presented in Sec.~\ref{sec:me}. In that section, the validity of the rate equations is confirmed
by the master equation for the reduced density matrix, which also enables us to study pure dephasing.
We summarize our work in Sec.~\ref{sec:summ}.
We use units $\hbar = k_B = 1$ throughout the paper.

\section{Model}
\label{sec:model}

The effective Hamiltonian $\hat{H}$ for a transmon qubit
can be split into three parts,
\be\label{Heff}
\hat{H} = \hat{H}_{\varphi} +\hat{H}_\qp + \delta\hat{H}\, ,
\ee
where the Hamiltonian $\hat{H}_{\varphi}$ describes the qubit,
$\hat{H}_\qp$ is the quasiparticle Hamiltonian, and $\delta\hat{H}$ the qubit-quasiparticles interaction term.
When restricted to the four lowest energy levels, the qubit Hamiltonian takes the form:
\be\label{Hqc}
\hat{H}_{\varphi} =  \, \frac{\EZO}{2} \hat\sigma^z - \frac{1+\hat\sigma^z}{2} \frac{\tilde\varepsilon_1(n_g)}{2} \hat\tau^z + \frac{1-\hat\sigma^z}{2}
\frac{\tilde\varepsilon_0(n_g)}{2} \hat\tau^z
\ee
where the coefficients $\EZO$ and $\tilde\varepsilon_{0,1}$ characterize the qubit spectrum [see Fig.~\ref{figlev}c)], including its dependence on background charges (and/or gates) via the dimensionless
voltage $n_g$. The (bare\cite{parren}) values of these coefficients are determined by the Josephson and charging energy $E_J$ and $E_C$, and for $E_J/E_C \gg 1$
they are given by\cite{transmon}
\bea
\EZO &=&  \omega_p - E_C\\
\tilde\varepsilon_i (n_g) &=& \varepsilon_i \cos \left(2\pi n_g\right) \\
\varepsilon_i &=& 4\omega_p (-1)^i \sqrt{\frac{2}{\pi}} \frac{2^{2i}}{i!}\left(\frac{8E_J}{E_C}\right)^{\frac{2i+1}{4}} e^{-\sqrt{8E_J/E_C}} \, , \quad \label{varepsid}
\eea
where the plasma frequency is
\be
\omega_p = \sqrt{8E_C E_J}\, .
\ee
The Pauli matrices $\hat\sigma^\mu$ act in the qubit level space (\textit{i.e.}, ground/excited state), while Pauli matrices $\hat\tau^\mu$ in the parity (even/odd) space.

The quasiparticle Hamiltonian is given by
\be\label{Hqp}
\hat{H}_\qp =\sum_{j=L,R} \hat{H}_{\qp}^j \, , \quad \hat{H}_{\qp}^j =
\sum_{a,\sigma} \epsilon_{a}^j
\hat\alpha^{j\dagger}_{a\sigma} \hat\alpha^j_{a\sigma},
\ee
where $\hat\alpha^j_{a\sigma}$($\hat\alpha^{j\dagger}_{a\sigma}$) are annihilation (creation)
operators for quasiparticles with spin $\sigma=\uparrow, \downarrow$ in electrode $j=L,R$ to
the left or right of the junction.
We assume for simplicity identical densities of states per spin direction $\nu_0$ and
the same superconducting gap $\Delta$ in both electrodes. The
quasiparticle energies are $\epsilon^j_{a} =
\sqrt{(\xi_{a}^j)^2+\Delta^2}$, with $\xi_{a}^j$
single-particle energy level $a$ in the normal state of electrode $j$.
The occupation probabilities of these levels are given by the distribution functions
\be
f^j(\xi^j_{a})= \langle\!\langle \hat\alpha_{a\uparrow}^{j\dagger} \hat\alpha^j_{a\uparrow}
\rangle\!\rangle_\qp
= \langle\!\langle \hat\alpha_{a\downarrow}^{j\dagger} \hat\alpha^j_{a\downarrow}\rangle\!\rangle_\qp\, ,
\quad j=L,R\, ,
\ee
where double angular brackets
$\langle\!\langle \ldots \rangle\!\rangle_\qp$ denote averaging over quasiparticle states.
We take the distribution functions to be independent of spin and equal in the two electrodes. We also
assume that $\delta E$, the characteristic energy of the quasiparticles above the gap, is small compared
to the gap, $\delta E \ll \Delta$, but
the distribution function is otherwise generic, thus allowing for non-equilibrium conditions.

The qubit-quasiparticle interaction term $\delta\hat{H}$ in \eref{Heff} accounts for tunneling and is discussed in detail in \ocite{prb}.
For our purposes, it can be written as [see Appendix~\ref{app:me}]
\be\label{HT}\begin{split}
\delta\hat{H} = \tilde{t} \sum_{a,b,\sigma}\bigg[
\left(c_{1} \frac{1+\hat\sigma^z}{2} + c_{0} \frac{1-\hat\sigma^z}{2}\right)\left(u^L_a u^R_b - v^L_a v^R_b \right)
\\ + i s \left(\hat\sigma^+ + \hat\sigma^- \right)\left(u^L_a u^R_b + v^L_a v^R_b\right)
 \bigg]\hat\tau^x
\hat\alpha^{L\dagger}_{a\sigma} \hat\alpha^{R}_{b\sigma}
+ \text{H.c.} \, ,
\end{split}\ee
where $\tilde{t}$ is the tunneling amplitude, the Bogoliubov amplitudes $u^j_{a}$, $v^j_{a}$ are real quantities, and the Pauli matrix $\hat\tau^x=\hat\tau^+ + \hat\tau^-$ accounts for the fact
that any time a single excitation tunneling event takes place, the qubit parity changes. (In contrast, pair tunneling does not affect
parity.) The coefficients $c_i$ and $s$ denote combinations of matrix elements for the operators
associated with the transfer of a single charge across the junction; for large ratio $E_J/E_C$,
they are given by [see Appendix~\ref{app:me}]
\bea\label{sdef}
s_{\phantom{i}} &=& \left(\frac{E_C}{8E_J}\right)^{1/4} \\
\label{cidef}
c_i & = & 1 - \left(i+\frac12\right)\sqrt{\frac{E_C}{8E_J}} - \frac32 \left(i+\frac14 \right)\frac{E_C}{8E_J} \, .
\eea

\subsection{Density matrix}
\label{sec:tme}

The total density matrix $\hat\rho_{tot}$ contains information about the qubit and quasiparticles. Since we are interested
in studying the dynamics of the qubit only, we will consider the reduced density matrix $\hat\rho$ obtained by
tracing out the quasiparticle degrees of freedom, $\hat\rho = \Tr_{\qp} \hat\rho_{tot}$. An eigenstate of the qubit is specified by a vector $|i, \alpha \rangle$, where $i=0$, $1$ denotes the qubit being in the ground or excited state, respectively, and $\alpha=e,\, o$ its even/odd parity.
Then in matrix form, the density matrix has four indices:
$\alpha$, $\beta$ for parity, and $i$, $j$ for state.
For the diagonal elements, we use the following decomposition in terms of Pauli matrices $\hat\sigma^\mu$ in the qubit state space and $\hat\tau^\mu$ in the parity space:
\bea
\rho_z & = & \Tr \left[\hat\rho \hat\sigma^z \right] \\
\rho_{1(0)}^z & = & \Tr \left[\hat\rho \frac{\hat1 \pm \hat\sigma^z}{2} \hat\tau^z \right] \, .
\eea
In this representation, $\rho_z$ is the occupation probability difference between the qubit levels after tracing out parity.
For the off-diagonal elements of $\hat\rho$, we find it convenient to distinguish terms with fixed parity or fixed qubit state as follows:
\bea
\rho_+^{e(o)} & = & \Tr \left[\hat\rho \hat\sigma^+ \frac{\hat1\pm\hat\tau^z}{2} \right] \label{r+eo}\\
\rho_{1(0)}^{+} & = & \Tr \left[\hat\rho \frac{\hat1\pm\hat\sigma^z}{2} \hat\tau^+ \right] \, .
\eea
The remaining elements are
\bea
\rho_+^+ & = & \Tr \left[\hat\rho \hat\sigma^+ \hat\tau^+ \right] \\
\rho_+^- & = & \Tr \left[\hat\rho \hat\sigma^+ \hat\tau^- \right] \, . \label{r+-}
\eea

Before considering the microscopic description of the qubit dynamics afforded by the reduced density matrix, we present briefly in the next section
phenomenological rate equations for the occupation probabilities of the four qubit states. The validity of these equations will then be confirmed when we turn to
the master equation for the reduced density matrix in Sec.~\ref{sec:me}.

\section{Rate equations}
\label{sec:re}

From a phenomenological point of view, it is straightforward to write down the most general system of rate equations that govern the time evolution of the occupation probability $P^\alpha_i(t)$ for state at level $i\in \{0,1\}$ with parity $\alpha \in \{e,o\}$:
\be\label{re}\begin{split}
\dot{P}^\alpha_i = & -\left(\Gamma^{\alpha\bar\alpha}_{i\bar{i}} + \Gamma^{\alpha\bar\alpha}_{ii} +\Gamma^{\alpha\alpha}_{i\bar{i}} \right) P^\alpha_i \\ & + \Gamma^{\bar\alpha \alpha}_{\bar{i} i} P^{\bar\alpha}_{\bar{i}} + \Gamma^{\bar\alpha \alpha}_{ii} P^{\bar\alpha}_i +\Gamma^{\alpha\alpha}_{\bar{i} i} P^{\alpha}_{\bar{i}}
\, .
\end{split}\ee
Here the dot represent differentiation with respect to time and we use the notation $\bar{i}=(i+1)$~mod~$2$ and $\bar{e} =o$.
The first term on the right hand side of \eref{re} accounts for the decrease in occupation due to events that change both parity and level (with rate $\Gamma^{\alpha\bar\alpha}_{i\bar{i}}$), parity but not level ($\Gamma^{\alpha\bar\alpha}_{ii}$), and level but not parity ($\Gamma^{\alpha\alpha}_{i\bar{i}}$). The last three terms account for the reverse processes.
The interaction with quasiparticles is responsible for the parity-changing events; the corresponding rates and their temperature dependence will be discussed in the next section. In contrast, to induce parity-preserving transitions a different mechanism must be at work, such as interaction with the noisy electromagnetic environment or surface impurities. While we will not explore these mechanisms here, we include their effects at this phenomenological level to enable comparison with experiments, in which a roughly temperature-independent, parity-preserving decay rate is measured.\cite{riste}

In principle one can obtain a full solution to the system in \eref{re} for arbitrary rates. However, we make the simplifying assumption that the rates are insensitive to the parity of the initial state, $\Gamma^{eo}_{ij} = \Gamma^{oe}_{ij}$ and $\Gamma^{ee}_{ij} = \Gamma^{oo}_{ij}$. For the parity-preserving rates, their (near) equality can be justified by observing\cite{ithier} that the rate is proportional to the spectral density $S(\omega)$ of the noise at the frequency given by the energy difference between levels; since the even and odd levels have almost the same energy differences, we can expect the rates to be the same up to small corrections.\cite{footbw} We will consider the validity of our simplifying assumption for the parity-changing rates in Sec.~\ref{sec:me}, where microscopic formulas for the rates are discussed.

To take advantage of the above assumption, and to facilitate comparison with the density matrix approach of the next section, we now introduce certain combinations of
occupation probabilities. The total probability
\be
P_0 = \sum_{i,\alpha} P^\alpha_i
\ee
is of course conserved, $\dot{P}_0 = 0$, as follows from \eref{re}, and is normalized to unity, $P_0=1$.
The difference in occupation probabilities between levels (irrespective of parity) is given by
\be
P_z = \left(P_1^e+ P_1^o\right) - \left(P_0^e + P_0^o\right) \, .
\ee
Thanks to our simplifying assumption, it obeys a simple equation
\be\label{Pz0eq}
\dot{P}_z = -\frac{1}{T_1} P_z + \Gamma^{ee}_{01} + \Gamma^{eo}_{01} - \Gamma^{ee}_{10} - \Gamma^{eo}_{10} \, ,
\ee
\be\label{t1def}
\frac{1}{T_1} = \Gamma^{ee}_{01} + \Gamma^{eo}_{01} + \Gamma^{ee}_{10} + \Gamma^{eo}_{10} \, ,
\ee
governing its relaxation to the steady state $P_{z,s} = T_1\left(\Gamma^{ee}_{01} + \Gamma^{eo}_{01} - \Gamma^{ee}_{10} - \Gamma^{eo}_{10}\right)$ with rate
$1/T_1$:
\be
P_z(t) = P_z(0) e^{-t/T_1} + P_{z,s} \left(1-e^{-t/T_1}\right) \, .
\ee

Two other probability differences are those for parity occupation at each qubit level:
\be
P_i^z = P_i^e - P_i^o \, , \qquad i\in\{0,1\} \, .
\ee
They obey coupled equations
\be\label{Pizeq}
\dot{P}_i^z = -2\Gamma_{ii}^{eo}P_i^z -\Gamma_{i\bar{i}}^{eo}P_i^z - \Gamma_{\bar{i}i}^{eo}P_{\bar{i}}^z -\Gamma_{i\bar{i}}^{ee}P_{i}^z
+\Gamma_{\bar{i}i}^{ee}P_{\bar{i}}^z
\ee
whose terms have simple interpretations: the first term on the right hand side represents a $T_1$-like, intralevel relaxation of parity, with the factor of 2 due to the assumed even/odd symmetry. The second and fourth terms are ``outgoing'' contributions from one level to the other; both parity-changing and -preserving processes decrease $P_i^z$. The third and last terms are ``incoming'' contributions from the other level; in this case, the parity-preserving process increases $P_i^z$, while the parity-changing ones
have the opposite effect.

The above probabilities can be combined into the parity autocorrelation function\cite{riste} $R_{ij}(t)$, which gives the correlation between initial and final parity knowing
that the qubit was initially prepared (finally measured) in state $i$ ($j$):
\be
R_{ij}(t) = \frac{P_i^z(0)P_j^z(t)}{\frac{1-(-1)^j P^0_z(t)}{2}} \, .
\ee
The knowledge of the initial qubit states translate into the initial conditions
\bea
P_z^0(0) &=& (-1)^{i+1} \\
P_{\bar{i}}^z(0) &=& 0 \, .
\eea
Usually the qubit excitation rates are much smaller than the corresponding decay rates, $\Gamma_{01}^{\alpha\beta} \ll \Gamma_{10}^{\alpha\beta}$; hence a reasonable approximation is to set $\Gamma_{01}^{\alpha\beta}$ to zero. Then solving the rate equations with the above initial conditions we find for the parity autocorrelation function:
\begin{subequations}\label{paf}
\bea
R_{00}(t) &=& \left[P_0^z(0)\right]^2 e^{-2\Gamma_{00}^{eo} t} \\
R_{11}(t) &=& \left[P_1^z(0)\right]^2 e^{-2\Gamma_{11}^{eo} t} \\
R_{10}(t) &=& \left[P_1^z(0)\right]^2\frac{\Gamma_{10}^{ee}-\Gamma_{10}^{eo}}{2\Gamma_{00}^{eo}-2\Gamma_{11}^{eo}-\Gamma_{10}^{ee}-\Gamma_{10}^{eo}} \times \\ &&
\frac{e^{-\left(2\Gamma_{11}^{eo}+\Gamma_{10}^{ee}+\Gamma_{10}^{eo}\right)t} - e^{-2\Gamma_{00}^{eo} t}}{1-e^{-\left(\Gamma_{10}^{ee}+\Gamma_{10}^{eo}\right)t}} \nonumber
\eea
while $R_{01}(t) = 0$ due to the assumption $\Gamma_{01}^{\alpha\beta} =0$.
In agreement with \ocite{riste}, we find that when the qubit is initially prepared in an eigenstate, $R_{10}(t\to 0) = \left(\Gamma_{10}^{ee}-\Gamma_{10}^{eo}\right)/\left(\Gamma_{10}^{ee}+\Gamma_{10}^{eo}\right)$.
Together with an independent determination of $T_1$, measurements of the three correlation functions in \eref{paf} give all the information needed to estimate the four rates $\Gamma_{00}^{eo}$, $\Gamma_{11}^{eo}$, $\Gamma_{10}^{ee}$, and $\Gamma_{10}^{eo}$. This procedure has indeed been employed successfully to measure the rates in \ocite{riste}. If the excitation rates $\Gamma_{01}^{ee}$ and $\Gamma_{01}^{eo}$ cannot be neglected, one needs to measure two more independent quantities, \textit{e.g.} the steady-state population difference $P_z^0(t\gg T_1)$ and the parity autocorrelation $R_{01}(t\to 0)$, and to modify the expressions in \eref{paf} to account for the finite excitation rates. Interestingly, the sign of $R_{01}(t\to 0)\propto\left(\Gamma_{01}^{ee}-\Gamma_{01}^{eo}\right)/\left(\Gamma_{01}^{ee}+\Gamma_{01}^{eo}\right)$ would give indication as to wether  ``hot'' quasiparticles are the main culprit for the finite steady-state qubit excitation, if $R_{01}(0)< 0$, or if some other parity-conserving mechanism is responsible, if $R_{01}(0)> 0$ (while in equilibrium $R_{01}(0)$ and $R_{10}(0)$ are proportional to each other and hence have the same sign, this is not necessarily true in non-equilibrium). We do not pursue this further here, but rather move on to
the microscopic validation of the rate equations by considering, in the next section, the master equation for the reduced density matrix.
\end{subequations}

\section{Master equation}
\label{sec:me}

The master equation governing the time evolution of the reduced density matrix $\hat\rho$ can be derived starting from the microscopic Hamiltonian \eref{Heff} and using
well-established approximation schemes (i.e., Born-Markov and rotating wave). This procedure is detailed in \ocite{prb2} and summarized in Appendix~\ref{app:sme} -- here
we present only the final expressions, starting with the equations for the diagonal components.

\subsection{Relaxation}

To evaluate the qubit relaxation rate, we consider the evolution equation for the qubit level occupation $\rho_z$:
\be\label{rz0eq}\begin{split}
\frac{d\rho_z}{dt} = \, & - \frac{1}{T_1}\rho_z
 -\frac12\left(\Gamma_{10}^{eo}+\Gamma_{10}^{oe}- \Gamma_{01}^{eo}-\Gamma_{01}^{oe}\right) 
\\ & - \left(\Gamma_{10}^{eo} - \Gamma_{10}^{oe}\right) \rho_1^z  + \left(\Gamma_{01}^{eo} - \Gamma_{01}^{oe}\right) \rho_0^z \, ,
\end{split}\ee
where
\be\label{t1def2}
\frac{1}{T_1} = \frac12\left(\Gamma_{10}^{eo}+\Gamma_{10}^{oe}+ \Gamma_{01}^{eo}+\Gamma_{01}^{oe}\right)
\ee
and the transition rates are\cite{prl,prb,prb2}
\be\label{decayrate}\begin{split}
\Gamma_{10}^{\alpha\beta} = \frac{16 E_J}{\pi \Delta} s^2 \int_\Delta^{+\infty}\!\!d\epsilon \, f(\epsilon)
\left[1-f\left(\epsilon+\omega^{\alpha\beta}\right)\right]
\\ \frac{\epsilon\left(\epsilon+\omega^{\alpha\beta}\right)+\Delta^2}
{\sqrt{\epsilon^2-\Delta^2}\sqrt{\left(\epsilon+\omega^{\alpha\beta}\right)^2-\Delta^2}}
\end{split}\ee
with
\be
\omega^{\alpha\beta} = \EZO - \mathcal{P}^\alpha \frac{\tilde\varepsilon_1}{2}- \mathcal{P}^\beta \frac{\tilde\varepsilon_0}{2} \, .
\ee
The parities are defined as $\mathcal{P}^e = 1$ and $\mathcal{P}^o = -1$. The $0\to 1$ rates are obtained by replacing $f \to (1-f)$ in \eref{decayrate}. When the characteristic quasiparticle energy is small compared to the qubit frequency, $\delta E \ll \EZO$, \eref{decayrate} gives a rate proportional to the quasiparticle density.\cite{prl,prb}. Here we note that for a quasi-equilibrium distribution function characterized by effective quasiparticle temperature $T_{e}$ and chemical potential $\mu_e$,
\be\label{qed}
f(\epsilon) = \frac{1}{e^{(\epsilon - \mu_e)/T_e}+1} \, ,
\ee
in the non-degenerate case $e^{-(\Delta - \mu_e)/T_e} \ll 1$ a good approximation for the integral in the right hand side of \eref{decayrate} is
\be\label{ia1}\begin{split}
\int_\Delta^{+\infty}\!\!d\epsilon \, f(\epsilon)
\left[1-f\left(\epsilon+\omega\right)\right]
\frac{\epsilon\left(\epsilon+\omega\right)+\Delta^2}
{\sqrt{\epsilon^2-\Delta^2}\sqrt{\left(\epsilon+\omega\right)^2-\Delta^2}}\simeq \\
\Delta e^{-(\Delta-\mu_e)/T_e} e^{\omega/2T_e} \left[K_0\left(\frac{\omega}{2T_e}\right)+\frac{\omega}{4\Delta}K_1\left(\frac{\omega}{2T_e}\right)\right] \, ,
\end{split}\ee
where $K_i$ denotes the modified Bessel function of the second kind.
For $T_e/\Delta \lesssim 0.2$ and $\omega/\Delta \lesssim 0.3$ the right hand side of \eref{ia1} deviates from the exact expression by less than 1\%.

Equation \rref{rz0eq} is the generalization of \eref{Pz0eq} to unequal even/odd rates (we remind that since we are considering only quasiparticle effects, the
are no parity-preserving transitions, $\Gamma^{\alpha\alpha}_{ij}=0$). Moreover, from the formula in \eref{decayrate} we can estimate the deviation from the
even/odd symmetry: assuming that quasiparticles are non-degenerate, $f(\epsilon) \ll 1$, we find
$|\left(\Gamma_{10}^{eo}-\Gamma_{10}^{oe}\right)|/\left(\Gamma_{10}^{eo}+\Gamma_{10}^{oe}\right) \lesssim |\tilde\varepsilon_1|/4\EZO$, and the inequality is saturated in the
case of qubit frequency large compared to quasiparticle energy above the gap, $\EZO \gg \delta E$.
Note that
already at moderate ratio $E_J/E_C=20$ we have $|\varepsilon_1|/4\EZO < 10^{-3}$, and that $|\varepsilon_1|/\EZO$ exponentially decreases as $E_J/E_C$ increases [cf. \eref{varepsid}]; therefore, the even/odd asymmetry in the relaxation rates is negligible.\cite{noteexc}

\subsection{Parity-switching rates}

The other two diagonal components of the density matrix also obey equations that generalize
\eref{Pizeq} to the case in which no even/odd symmetry for the rates is present:
\bea
\frac{d\rho_0^z}{dt} & = & -\left(\Gamma_{00}^{eo}+\Gamma_{00}^{oe}+\frac12\Gamma_{01}^{eo} +\frac12\Gamma_{01}^{oe} \right)\rho_0^z
\\ && -\frac12 \left(\Gamma_{10}^{eo} + \Gamma_{10}^{oe} \right)\rho_1^z \nonumber \\
&& +\frac14 \left[\Gamma_{01}^{eo}-\Gamma_{10}^{eo}+2\Gamma_{00}^{eo}- \left(e\leftrightarrow o\right) \right] \rho_z \nonumber
\\ && -\frac14 \left[\Gamma_{01}^{eo}+\Gamma_{10}^{eo}+2\Gamma_{00}^{eo}- \left(e\leftrightarrow o\right) \right] \nonumber
\eea
and
\bea
\frac{d\rho_1^z}{dt} & = & -\left(\Gamma_{11}^{eo}+\Gamma_{11}^{oe}+\frac12\Gamma_{10}^{eo} +\frac12\Gamma_{10}^{oe} \right)\rho_1^z
\\ & & -\frac12 \left(\Gamma_{01}^{eo} + \Gamma_{01}^{oe} \right)\rho_0^z \nonumber \\
&& +\frac14 \left[\Gamma_{01}^{eo}-\Gamma_{10}^{eo}-2\Gamma_{11}^{eo}- \left(e\leftrightarrow o\right) \right] \rho_z \nonumber
\\ && -\frac14 \left[\Gamma_{01}^{eo}+\Gamma_{10}^{eo}-2\Gamma_{11}^{eo}- \left(e\leftrightarrow o\right) \right] \, , \nonumber
\eea
where the parity switching rates are
\be\label{g00eo}\begin{split}
\Gamma_{00}^{eo} = \frac{16 E_J}{\pi \Delta} c_0^2 \int_\Delta^{+\infty}\!\!d\epsilon \, f(\epsilon)
\left[1-f\left(\epsilon+\tilde\varepsilon_0\right)\right]
\\ \frac{\epsilon\left(\epsilon+\tilde\varepsilon_0\right)-\Delta^2}
{\sqrt{\epsilon^2-\Delta^2}\sqrt{\left(\epsilon+\tilde\varepsilon_0\right)^2-\Delta^2}} \, ,
\end{split}\ee
\be\label{g11oe}\begin{split}
\Gamma_{11}^{oe} = \frac{16 E_J}{\pi \Delta} c_1^2 \int_\Delta^{+\infty}\!\!d\epsilon \, f(\epsilon)
\left[1-f\left(\epsilon+\tilde\varepsilon_1\right)\right]
\\ \frac{\epsilon\left(\epsilon+\tilde\varepsilon_1\right)-\Delta^2}
{\sqrt{\epsilon^2-\Delta^2}\sqrt{\left(\epsilon+\tilde\varepsilon_1\right)^2-\Delta^2}} \, ,
\end{split}\ee
and the rates with even/odd exchanged are obtained by the replacement $f \to (1-f)$.

As discussed above, for the qubit transition the deviations from even/odd symmetry are small in the parameter $|\tilde\varepsilon_1|/\EZO$, which depends solely on the qubit properties; the only assumption needed for quasiparticles is that they are non-degenerate. In contrast, for the parity-switching rates we must compare $\tilde\varepsilon_i$
to the characteristic quasiparticle energy $\delta E$: if $|\tilde\varepsilon_i| \gg \delta E$, it means that there are no quasiparticle with sufficient energy
to excite the qubit; hence, in this case we have $\Gamma_{00}^{eo} \gg \Gamma_{00}^{oe}$ and $\Gamma_{11}^{oe} \gg \Gamma_{11}^{eo}$. In practice, however, the quasiparticle energy is at least of order of the base temperature (so larger than 10~mK, or 200~MHz); since for $E_J/E_C >20$ we have $|\varepsilon_1| \lesssim 10^{-3} \EZO$, for qubits with frequency in the 1-10~GHz range this implies $|\tilde\varepsilon_i| \ll \delta E$. In this regime of small splitting compared to $\delta E$ and for non-degenerate quasiparticles, using \esref{g00eo}-\rref{g11oe} we estimate $|\Gamma_{ii}^{eo}-\Gamma_{ii}^{oe}|/(\Gamma_{ii}^{eo}+\Gamma_{ii}^{oe}) \sim |\tilde\varepsilon_i|/\delta E \ll 1$ -- we find again that the assumption of even/odd symmetry for the rates is justified; hence, \esref{Pz0eq} and \rref{Pizeq} are indeed good approximations.
Note that for the quasi-equilibrium distribution in \eref{qed}, the estimate for the rate asymmetry follows directly from the detailed balance relation $\Gamma_{ii}^{oe}/\Gamma_{ii}^{eo}= e^{-\tilde\varepsilon_i/T_e}$. In the non-degenerate case, for $T_e/\Delta \lesssim 0.2$ and $\varepsilon/\Delta \lesssim 0.3$ an accurate approximate expression (relative error at most $\sim1\%$) for the integral in the right hand sides of \esref{g00eo}-\rref{g11oe} is
\be\begin{split}
\int_\Delta^{+\infty}\!\!d\epsilon \, f(\epsilon)
\left[1-f\left(\epsilon+\varepsilon\right)\right]
\frac{\epsilon\left(\epsilon+\varepsilon\right)-\Delta^2}
{\sqrt{\epsilon^2-\Delta^2}\sqrt{\left(\epsilon+\varepsilon\right)^2-\Delta^2}}\simeq \\
\frac{\omega}{2} e^{-(\Delta-\mu_e)/T_e} e^{\omega/2T_e} \left[K_1\left(\frac{\omega}{2T_e}\right)-\frac{\omega}{4\Delta} K_0\left(\frac{\omega}{2T_e}\right)\right] \, .
\end{split}\ee

\begin{figure}[t]
 \includegraphics[width=0.45\textwidth]{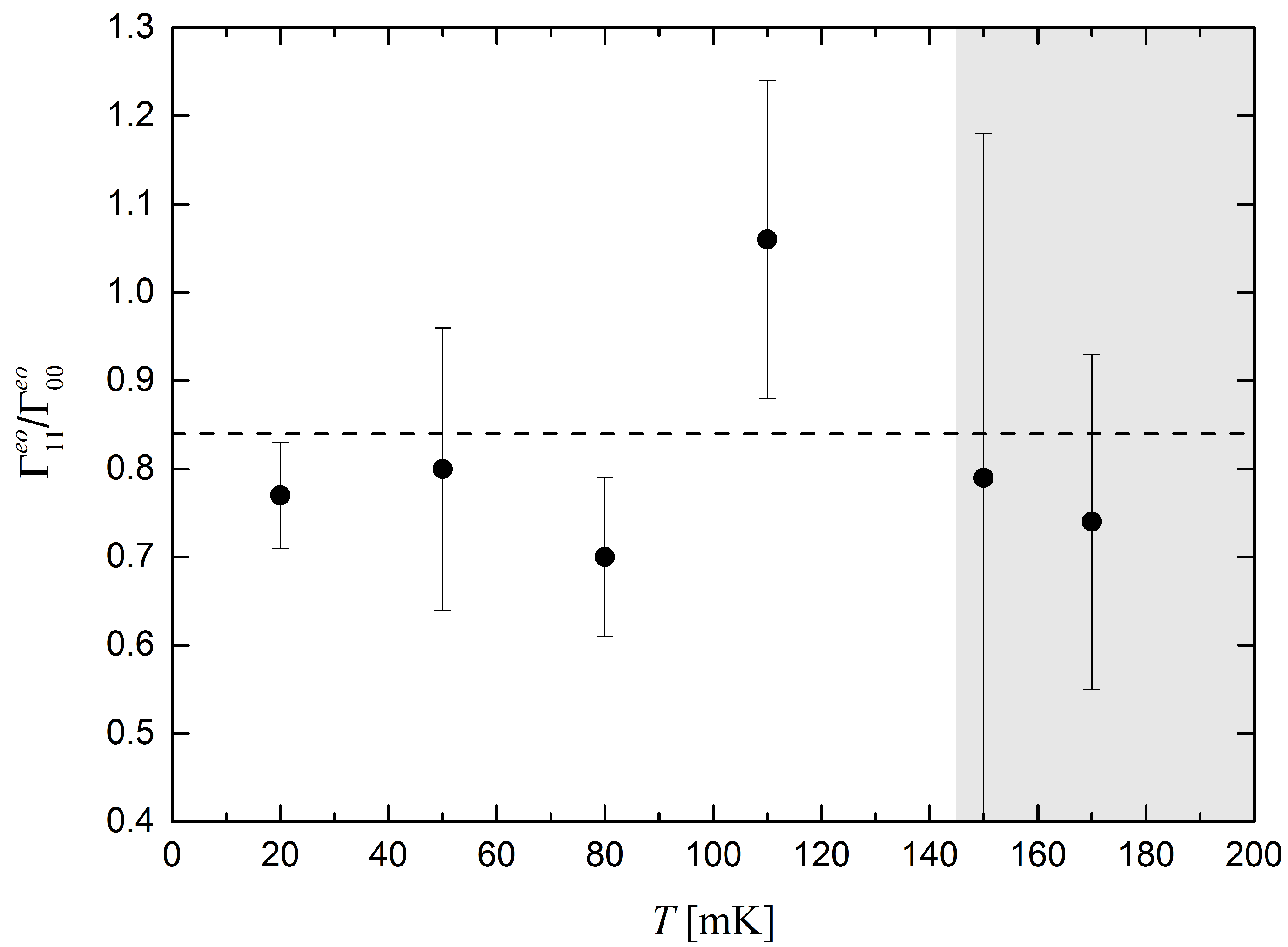}
 \caption{Points: experimental parity-switching rates ratio obtained from the measurements of the rates in \ocite{riste}. Dashed horizontal line: theoretical prediction from \eref{geor}. The shaded
 region at higher temperature denotes the (experimentally determined) regime of thermal equilibrium.}
 \label{figeo}
\end{figure}

When the condition $|\tilde\varepsilon_i| \ll \delta E$ is satisfied, the formulas for the parity switching rates simplify to
\be\label{psrs}
\Gamma_{ii}^{eo} \simeq \Gamma_{ii}^{oe} \approx \frac{16 E_J}{\pi \Delta} c_i^2 \int_\Delta^{+\infty}\!\!d\epsilon \, f(\epsilon)
\left[1-f\left(\epsilon\right)\right] \, .
\ee
Then, independent of the specific form of the quasiparticle distribution function, the ratio between the parity switching rates of the two levels depends solely on the matrix elements $c_i$:
\be\label{geor}
\frac{\Gamma_{11}^{oe}}{\Gamma_{00}^{eo}} \simeq \left(\frac{c_1}{c_0}\right)^2 \approx 1 - 2\sqrt{\frac{E_C}{8E_J}} - 3 \frac{E_C}{8E_J} < 1 \, .
\ee
In Fig.~\ref{figeo} we compare the ratio given by \eref{geor} with that extracted form the experimental data in \ocite{riste}; it was found there that at sufficiently high temperature (shaded area) the data are close to the thermal equilibrium expectation, but that large deviations are present at lower temperatures. Nonetheless, within experimental errors the ratio between the parity switching rates is found to be roughly constant, and consistent with \eref{geor}, both in and out of equilibrium.

We can glean some information on the quasiparticle distribution by comparing the parity-switching rates to the quasi\-particle-induced decay rate $\Gamma_{10}^{eo}$. In quasi-equilibrium [\eref{qed}], for non-degenerate quasiparticles their ratio is
\be\label{psrera}
\frac{\Gamma_{ii}^{eo}}{\Gamma_{10}^{eo}} \simeq \frac{c_i^2}{s^2}\sqrt{\frac{T_e \EZO}{\pi \Delta^2}}
\ee
for $\EZO\ll \Delta$. Note that the ratio of matrix elements [\esref{sdef}-\rref{cidef}] in the first factor on the right hand side is large in the parameter $E_J/E_C$ and can compensate for the smallness of the square root term. Indeed, for aluminum qubits ($\Delta\sim 2.2$~K) this ratio ranges from about 0.1 (at $T_e=20$~mK, $\EZO = 1$~GHz, and $E_J/E_C =20$) to about 2
($T_e=200$~mK, $\EZO = 10$~GHz, and $E_J/E_C =80$), thus predicting that parity switching and relaxation rates are within one order of magnitude from each other; this
is qualitatively consistent with measurements.~\cite{riste} However, \eref{psrera} also predicts that as temperature is lowered, the parity switching time should become longer compared to the relaxation time; this is in contrast with the experimental observation that the parity switching time is longer than the relaxation time at the highest measured temperature, but shorter at the lowest temperature. Thus, the measurements seem to indicate that there are deviations from the quasi-equilibrium assumption.

\subsection{Dephasing}
\label{sec:deph}

In a two-level system, the dephasing rate determines the time decay of the single off-diagonal element $\rho_+$ of the density matrix. For the trasmon, due to presence of 4 levels, there are 6 off-diagonal elements, as defined in \esref{r+eo}-\rref{r+-}. Of those elements, $\rho_+^{e}$ and $\rho_+^{o}$ describe superpositions of qubit states with a given parity, and their sum the qubit coherence after tracing out parity. Here we are indeed interested in the coherence of qubit states (rather than among parity states in a given qubit level); thus, we focus on $\rho_+^{e(o)}$ only. They obey the coupled equations:\cite{freqshift}
\bea\label{rpee}
\frac{d\rho_+^e}{dt} & = & i\left(\EZO- \frac{\tilde\varepsilon_1+\tilde\varepsilon_0}{2}\right) \rho_+^e-\frac12 \left(\Gamma_{10}^{eo} + \Gamma_{01}^{eo}\right)\rho_+^e \\
&& -\frac14 \left(\Gamma_{00}^{eo}+\Gamma_{00}^{oe}+ \Gamma_{11}^{eo}+ \Gamma_{11}^{oe}\right) \rho_+^e \nonumber \\
&& +\frac14\left(\frac{c_1}{c_0}\Gamma_{00}^{eo}+\frac{c_1}{c_0}\Gamma_{00}^{oe}+ \frac{c_0}{c_1}\Gamma_{11}^{eo}+ \frac{c_0}{c_1}\Gamma_{11}^{oe}\right) \rho_+^o \nonumber
\eea
and
\bea\label{rpoe}
\frac{d\rho_+^o}{dt} &= & i\left(\EZO+ \frac{\tilde\varepsilon_1+\tilde\varepsilon_0}{2}\right) \rho_+^o-\frac12 \left(\Gamma_{10}^{oe} + \Gamma_{01}^{oe}\right)\rho_+^o \\
&& -\frac14 \left(\Gamma_{00}^{eo}+\Gamma_{00}^{oe}+ \Gamma_{11}^{eo}+ \Gamma_{11}^{oe}\right) \rho_+^o \nonumber\\
&& +\frac14\left(\frac{c_1}{c_0}\Gamma_{00}^{eo}+\frac{c_1}{c_0}\Gamma_{00}^{oe}+ \frac{c_0}{c_1}\Gamma_{11}^{eo}+ \frac{c_0}{c_1}\Gamma_{11}^{oe}\right) \rho_+^e \, . \nonumber
\eea
In both equations, the last term of the first line describes decoherence due to relaxation; the last two lines account
for quasiparticle tunneling events which change parity but not qubit level.

In the practically relevant case of small level splitting compared to quasiparticle energy, $|\tilde\varepsilon_i| \ll \delta E$, the
approximations in \esref{psrs}-\rref{geor} lead to a simplified set of equations. Considering the linear combinations
$\rho_+ = \rho_+^e + \rho_+^o$ and $\rho_+^z = \rho_+^e - \rho_+^o$, the simplified equations read
\bea
\frac{d\rho_+}{dt} & = & i\EZO \rho_+ -\frac{1}{2T_1}\rho_+ -\frac12 \left(\frac{c_1}{c_0}-1\right)^2\Gamma_{00}^{eo} \rho_+
-i \bar\varepsilon \rho_+^z \nonumber \\ && \\
\frac{d\rho_+^z}{dt} & = & i\EZO \rho_+^z -\frac{1}{2T_1}\rho_+^z -\frac12 \left(\frac{c_1}{c_0}+1\right)^2\Gamma_{00}^{eo} \rho_+^z
-i\bar\varepsilon \rho_+ \nonumber \\ &&
\eea
where $\bar\varepsilon = (\tilde\varepsilon_1 + \tilde\varepsilon_0)/2$ and $T_1$ is defined as in \eref{t1def2}. If the terms proportional to $\bar\varepsilon$ can be neglected, the equations decouple and the (approximate) solution for $\rho_+$, describing the qubit decoherence, is
\be
\rho_+(t) = \rho_+(0) e^{i\EZO t} e^{-\left(1/2T_1 + \Gamma_\phi \right)t} \, ,
\ee
with\cite{prb2}
\be
\Gamma_\phi = \frac12 \left(\frac{c_1}{c_0}-1\right)^2\Gamma_{00}^{eo} \, .
\ee
To see when neglecting $\bar\varepsilon$ is justified, consider the general solution for $\rho_+$:
\be
\rho_+(t) = r_+ e^{\lambda_+ t} + r_- e^{\lambda_- t} \, ,
\ee
where the coefficients $r_\pm$ are determined by the initial conditions,
\be\label{lpm}
\lambda_{\pm} = i\EZO -\frac{1}{2T_1} -\Gamma_p -\frac{\Gamma_\phi}{2} \pm \sqrt{\left(\Gamma_p -\frac{ \Gamma_\phi}{2}\right)^2 - \bar\varepsilon^2} \, ,
\ee
and
\be
\Gamma_p = \frac14 \left(\frac{c_1}{c_0}+1\right)^2\Gamma_{00}^{eo} \, .
\ee
The rate $\Gamma_p$ is, at leading order in the small parameter $E_C/E_J$, the parity switching rate. Moreover, we have
\be
\frac{\Gamma_\phi}{\Gamma_p} \simeq \frac{E_C}{16E_J} \ll 1 \, .
\ee
Note that since $0 > \Re \lambda_+ \ge \Re \lambda_-$, the decoherence rate is determined by $\lambda_+$.

Introducing as usual the decoherence and pure dephasing times $T_2$ and $T_\phi$ via
\be
\frac{1}{T_2} = -\Re \lambda_+ = \frac{1}{2T_1} + \frac{1}{T_\phi} \, ,
\ee
we can distinguish three regimes: in the limit of small splitting the dephasing rate
is
\be\label{pdss}
\frac{1}{T_\phi} \simeq \Gamma_\phi\, , \qquad \bar\varepsilon \ll \sqrt{2\Gamma_p \Gamma_\phi} \, .
\ee
This is the regime considered above in which $\bar\varepsilon$ can be neglected. Note that in this case we recover the pure
dephasing rate calculated for a two-level system in \ocite{prb2}; this is expected, since at sufficiently small splitting the different parities cannot be distinguished. However, as we show next, the two-level approximation does not apply anymore as the splitting increases.

\begin{figure}[t]
 \includegraphics[width=0.48\textwidth]{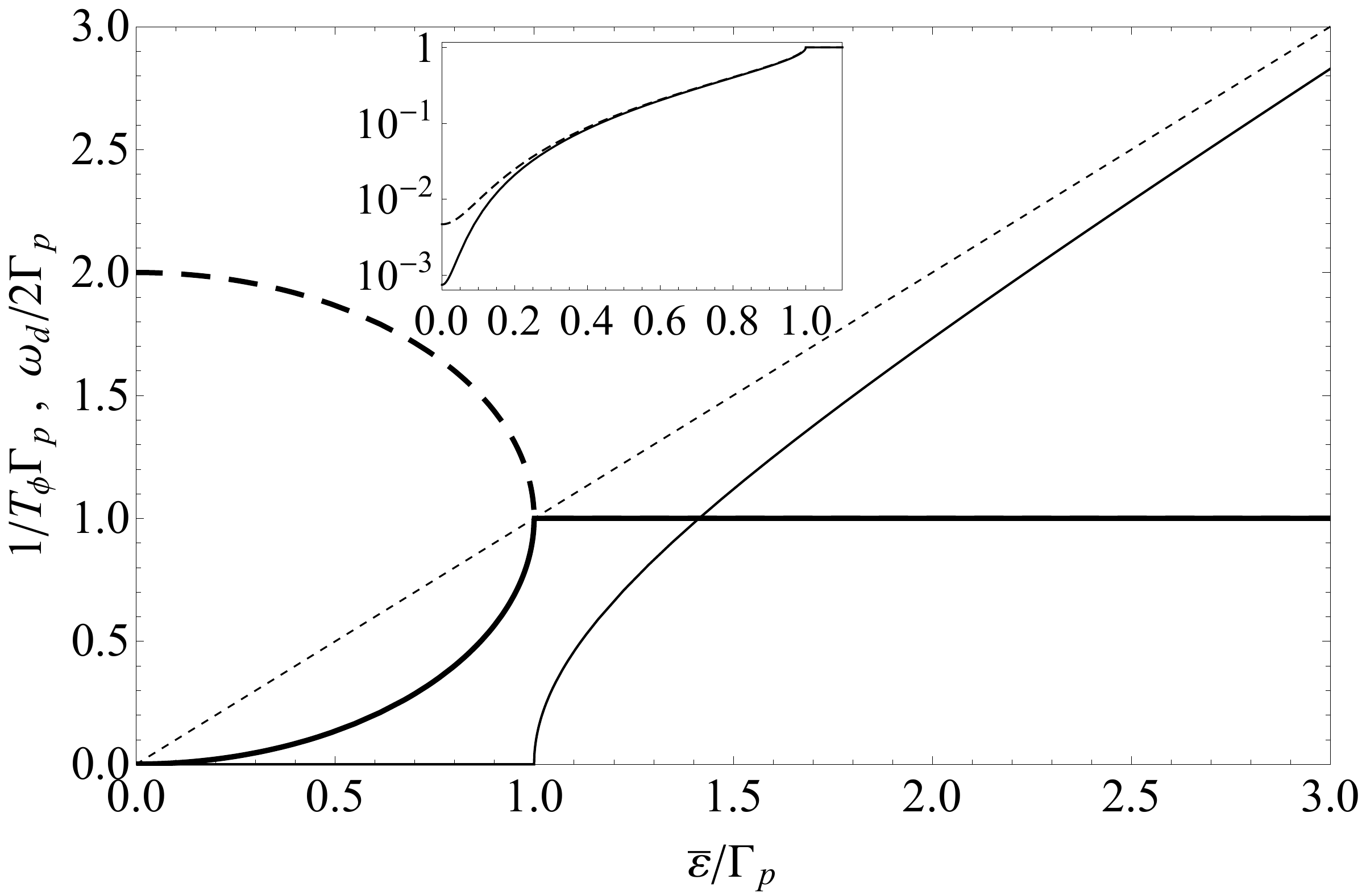}
 \caption{The thick solid line depicts the (normalized) dephasing rate $1/T_\phi \Gamma_p$ vs. the (normalized) splitting $\bar\varepsilon/\Gamma_p$, while the thick dashed line represent the faster dephasing rate (from $\Re\lambda_-$) of the other component of $\rho_+$; note that the two rates become equal at $\bar\varepsilon/\Gamma_p \simeq 1$. The thin solid line gives the (normalized) Ramsey frequency difference $\omega_d$ as function of the spectroscopic frequency difference $2\bar\varepsilon$; $\omega_d$ is always smaller than $2\bar\varepsilon$ [cf. dotted line]. Inset: $1/T_\phi \Gamma_p$ vs. $\bar\varepsilon/\Gamma_p$ for different values of $E_J/E_C$, namely 100 for the solid line and 20 for the dashed line: for larger $E_J/E_C$ the dephasing rate is smaller at a given splitting.}
 \label{figtpwd}
\end{figure}

At larger splitting, the behavior of the transmon resembles that of a qubit coupled to a two-level fluc\-tu\-ator:\cite{rtn1}
for intermediate splitting, the dephasing rate is quadratic in the splitting,
\be\label{pdis}
\frac{1}{T_\phi} \simeq \frac{\bar\epsilon^2}{2\Gamma_p} \, , \qquad \sqrt{2\Gamma_p \Gamma_\phi} \ll \bar\varepsilon \ll \Gamma_p \, .
\ee
At sufficiently large splitting, dephasing is determined by the parity switching rate
\be\label{pdls}
\frac{1}{T_\phi} \simeq \Gamma_p \, , \qquad \bar\varepsilon \gtrsim \Gamma_p
\ee
and $\rho_+$ is the sum of two terms with different frequencies, since $\Im \lambda_\pm \simeq \EZO \pm \bar\varepsilon$. These two frequencies can be seen in a Ramsey experiment\cite{riste} -- the Ramsey signal is the sum of two sinusoids with different frequencies but decaying at the same rate.
In Fig.~\ref{figtpwd} we show the variations of (normalized) dephasing rate $1/T_\phi$ and Ramsey fringes frequency difference $\omega_d = \Im \lambda_+ - \Im \lambda_-$ as functions of $\bar\varepsilon$, as obtained from \eref{lpm}. We note that the transition between the intermediate and large splitting regimes is sharp, as the
corresponding transition in the case of a qubit interacting with a two-level fluctuator, while the passage from intermediate to small splitting is a smooth cross-over.
Moreover, the frequency difference in a Ramsey experiment is always smaller than the spectroscopic frequency difference $2\bar\varepsilon$. The similarity between dephasing due to parity switching and the effect of a fluctuator indicates that the dephasing can be attributed to the change in qubit frequency after a parity-switching event; therefore the latter, in contrast to quasiparticle relaxation, does not destroy the superposition of qubit states.\cite{caseB}

\begin{figure}[!t]
 \includegraphics[width=0.48\textwidth]{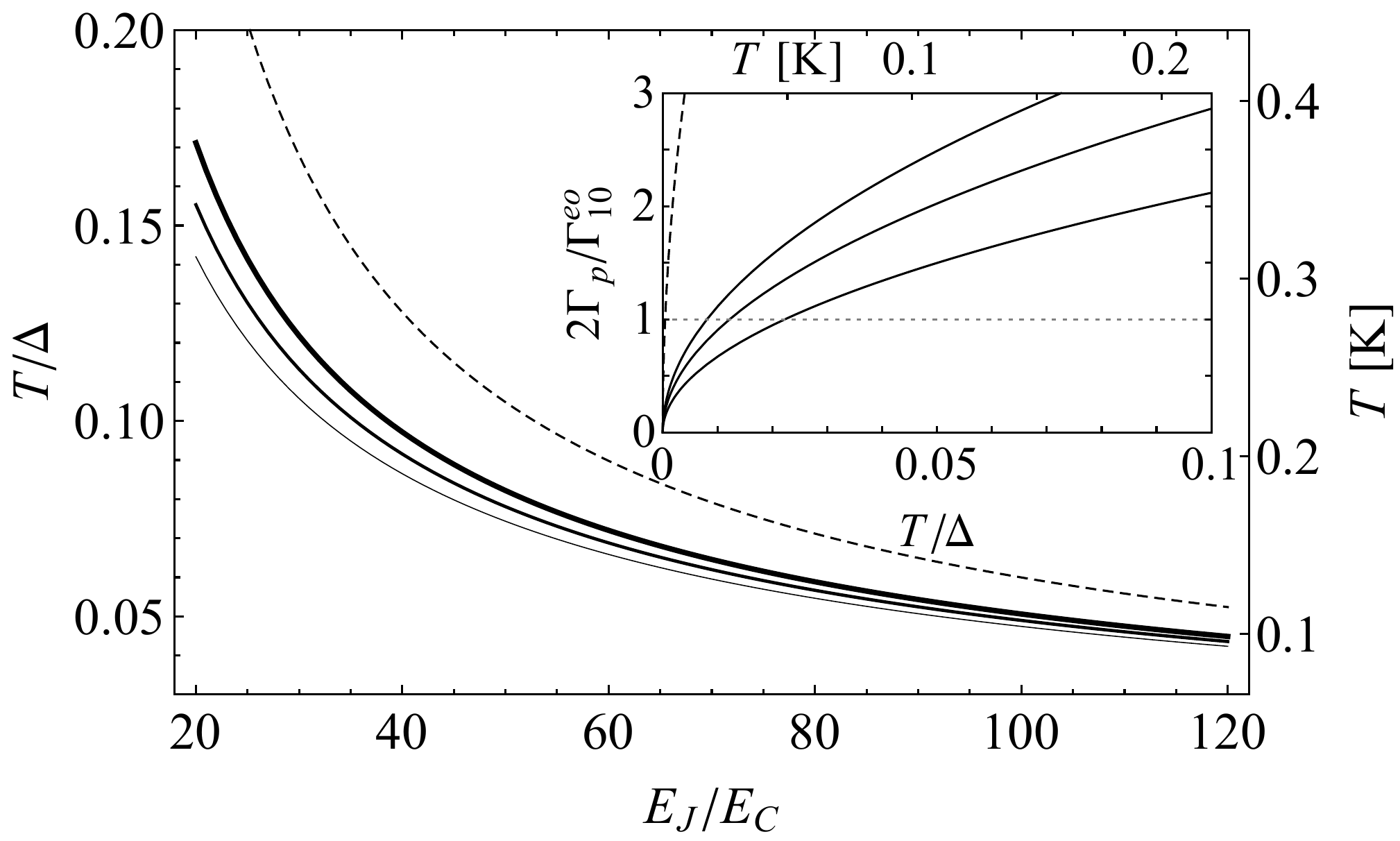}
 \caption{The thick solid line gives the points in the $E_J/E_C$-$T/\Delta$ plane where $\bar\varepsilon/\Gamma_p=1$; here thermal equilibrium and $n_g=1/2$ are assumed. The thinner (thinnest) solid line is where $\bar\varepsilon/\Gamma_p=2$ ($\bar\varepsilon/\Gamma_p=4$), and the dashed line where $\bar\varepsilon = \sqrt{2\Gamma_p\Gamma_\phi}$. Inset: solid lines show $\Gamma_p$ normalized by half the decay rate $\Gamma_{10}^{eo}$ as function of temperatures for (bottom to top) $E_J/E_C=20$, $50$, and $100$ in a small junction with $E_J/\Delta = 0.5$. Dashed line: normalized $\Gamma_p$ at $E_J/E_C = 20$ for a large junction with $E_J/\Delta = 25$. The absolute temperature scale on the right (top in the inset) is calculated for $\Delta =2.2$~K.}
 \label{fig3}
\end{figure}

We can summarize the above discussion as follows: in the regime of splitting large compared to parity switching rate, the latter determines the pure dephasing rate and
the Ramsey signal is the sum of two terms oscillating with different frequencies; in the opposite case of small splitting, the dephasing rate is suppressed below the parity switching rate and the Ramsey signal oscillates at the usual single frequency given by the detuning from the qubit frequency $\EZO$. To investigate which of these two situations is experimentally realized, in Fig.~\ref{fig3} we plot, assuming thermal equilibrium, the transitions temperatures between the three
regimes as function of $E_J/E_C$. The thick solid line indicates that below $\sim 100$~mK in aluminum qubits (right temperature scale) the splitting should be larger than the switching rate.
Therefore, at low temperatures
the pure dephasing rate should be determined by the parity switching rate and the latter, as shown in the inset, is generally of the order of or larger than the quasiparticle-induced $1/2T_1$ contribution to decoherence for small junctions ($E_J \lesssim \Delta$), while is generally much larger than
$1/2T_1$ for larger-area junctions with $E_J > \Delta$. Thus, at low temperatures parity switching could be a dominant source of dephasing in a single-junction transmon, especially for larger junction (if non-quasiparticle processes are not the factor limiting the coherence time; experimental evidence suggests that in current experiments photon shot noise is  a more important source of dephasing, see \ocite{psn}).

\section{Summary}
\label{sec:summ}

In this paper we have studied parity switching caused by quasiparticle tunneling in single-junction transmons. The parity-switching rates can be obtained from measurement of the parity autocorrelation function, see \ocite{riste} and Sec.~\ref{sec:re}. As we argue in Sec.~\ref{sec:me}, the experimentally relevant regime is that in which the splitting between the transmon states with different parities is small compared to the characteristic quasiparticle energy above the gap. In this regime, we find that the quasiparticle-induced relaxation and parity-switching rates are even/odd symmetric, \textit{i.e.}, they do not depend on the initial state parity.
Moreover, the ratio between parity-switching rates of different qubit levels does not depend on the quasiparticle distribution function, but only on the ratio between charging and Josephson energies, see \eref{geor}. This theoretical result is compared to experimental data in Fig.~\ref{figeo}, both in and out of equilibrium.

In Sec.~\ref{sec:deph} we have considered the role of parity switching in the transmon dephasing. We identify three regimes for the pure dephasing rate at different ratios of splitting $\bar\varepsilon$ to parity switching rate $\Gamma_p$, see \esref{pdss}-\rref{pdls}. In particular, for $\bar\varepsilon$ larger than $\Gamma_p$, the pure dephasing rate is given by $\Gamma_p$ -- as discussed in the text describing Fig.~\ref{fig3}, this regime is the relevant one when the system is cooled below about 100~mK. Based on the rates measured in \ocite{riste}, our results indicate that pure dephasing by quasiparticles could become a significant source of decoherence, if the coherence time of a transmon can be extended by another order of magnitude by suppressing other decoherence mechanisms.

\acknowledgments

Discussions with D. Rist\`e, L. DiCarlo, L. Glazman, R. Schoelkopf, and D. DiVincenzo are gratefully acknowledged. This work was supported in part by the EU under REA grant agreement CIG-618258.

\appendix

\section{Matrix elements and effective Hamiltonian}
\label{app:me}

In this appendix we briefly motivate the form of the qubit-quasiparticle interaction Hamiltonian $\delta\hat{H}$ in \eref{HT}. Our starting point is the quasiparticle
tunneling Hamiltonian\cite{prl} which can be written as
\be\label{HTo}\begin{split}
\hat{H}_T = \ & \tilde{t}\sum_{a,b,\sigma} \Big[ \cos\frac{\hat\varphi}{2} \left(u_a^Lu_b^R-v_a^Lv_b^R\right) + \\ & i \sin\frac{\hat\varphi}{2} \left(u_a^Lu_b^R+v_a^Lv_b^R\right)  \Big] \hat\alpha^{L\dagger}_{a\sigma} \hat\alpha^R_{b\sigma} + \mathrm{H.c.} \, .
\end{split}\ee
For quasiparticles with energy close to the gap, the combination $ \left(u_a^Lu_b^R-v_a^Lv_b^R\right)$ in the first term in square bracket is suppressed compared to  $\left(u_a^Lu_b^R+v_a^Lv_b^R\right)$ in the second one when
$\delta E$, $\EZO \ll 2\Delta$ -- that is why only the second term was
retained in \ocite{prb}. Here, as in \ocite{prb2}, we go beyond that approximation and consider the matrix elements of both $\sin\frac{\hat\varphi}{2}$ and
$\cos\frac{\hat\varphi}{2}$.

For the qubit wavefunctions, we can use the tight-binding form introduced in Appendix~B of \ocite{prb}. Then it is straightforward to show that matrix elements between states with the same parity vanish:
\be
\langle i, \alpha | \sin\frac{\hat\varphi}{2} | j, \alpha \rangle = \langle i, \alpha | \cos\frac{\hat\varphi}{2} | j, \alpha \rangle = 0 \, .
\ee
As for the matrix elements between states with different parity, for the operator $\sin\frac{\hat\varphi}{2}$ they where calculated in Appendices~B and E of \ocite{prb}:
\be\label{rmes}
\langle 1,\alpha | \sin\frac{\hat\varphi}{2} | 0, \bar\alpha \rangle \simeq \left(\frac{E_C}{8E_J}\right)^{1/4},
\ee
\be\label{psmes}\begin{split}
& \left| \langle i,\alpha | \sin\frac{\hat\varphi}{2} | i, \bar\alpha \rangle \right| \simeq \\ & \left| \sin \left(2\pi n_g\right) \right| \left(\frac23\right)^{2/3}
\Gamma\left(\frac13\right) \left(\frac{E_C}{8E_J}\right)^{1/6} \frac{\varepsilon_i}{\omega_p} \, ,
\end{split}\ee
with $\Gamma$ denoting the gamma function. Using the same approaches detailed in the above-mentioned appendices of \ocite{prb}, we find (for $i=0,1$)
\be\label{psmec}
\langle i,\alpha | \cos\frac{\hat\varphi}{2} | i, \bar\alpha \rangle \simeq 1 - \left(i+\frac12\right)\sqrt{\frac{E_C}{8E_J}} - \frac32 \left(i+\frac14 \right)\frac{E_C}{8E_J},
\ee
\be\label{rmec}
\left| \langle 1,\alpha | \cos\frac{\hat\varphi}{2} | 0, \bar\alpha \rangle \right| \propto \left| \cos \left(2\pi n_g\right) \right| \frac{\sqrt{|\varepsilon_0\varepsilon_1}|}{\omega_p} \left(\frac{E_C}{E_J}\right)^{1/3} \, .
\ee

Comparing \eref{rmec} and \eref{rmes}, it is clear why the former matrix element can always be neglected in comparison with the latter: as mentioned above, the combinations of Bogoliubov amplitudes in \eref{HTo} suppress the $\cos\hat\varphi/2$ contributions
in comparison to the $\sin\hat\varphi/2$ ones, and moreover for relaxation/excitation processes the matrix element of cosine
is exponentially smaller than that of sine. The situation is only slightly more complicated when considering the parity switching
matrix elements in \esref{psmes} and \rref{psmec}, since one has to allow for the possibility that the suppression in the Bogoliubov
amplitude combination of the cosine term could compensate for the exponential suppression of the sine term. We can see that this
possibility can always be neglected in practice by comparing the respective contributions to the parity switching rate. We consider for concreteness the experimentally relevant case of splitting small compared to effective temperature, $|\varepsilon_1| \ll T_e$ (for simplicity, we set $\mu_e=0$). Then
for the cosine contribution, the parity switching rate in the excited state is given by \eref{psrs}:
\be\label{cps}
\Gamma_{11}^{eo} \approx \frac{16E_J}{\pi} \frac{T_e}{\Delta} e^{-\Delta/T_e}
\ee
The sine contribution, denoted by $\Gamma_{e \to 0}^{(1)}$, is given in Eq.~(C8) of \ocite{prb}, and diverges for $n_g \to 1/4$ -- this divergence can in principle compensate for the exponential smallness of the sine matrix element. Parameterizing $n_g$ as
\be
n_g = \frac14 + \frac{\eta}{2\pi} \, ,
\ee
for $\eta \to 0$ we have
\be\label{sps}
\Gamma_{e \to 0}^{(1)} \approx \frac{16E_J}{\pi} e^{-\Delta/T_e} \left(\frac{E_C}{E_J}\right)^{1/3}\left(D \frac{\varepsilon_1}{\omega_p}\right)^2 \ln \frac1\eta \, .
\ee
Even choosing the most favorable realistic values of the parameters ($T_e/\Delta \sim 0.01$, $E_J/E_C \sim 20$), the rate in
\eref{sps} becomes comparable to that in \eref{cps} only for extremely small values of $\eta$, $\eta \sim 10^{-10^3}$. Therefore we
can in practice neglect the sine contribution to the parity switching rate.

Having discussed the various matrix elements in the preceding paragraphs, we can now project \eref{HTo} onto the four lowest level, and neglecting exponentially small terms [\eref{psmes} and \rref{rmec}] we arrive at \eref{HT}.

\section{Derivation of the master equation}
\label{app:sme}

The derivation of the master equation using the Hamiltonian in \eref{Heff} starts from the von Neumann equation and employs the Born-Markov and rotating wave approximations.\cite{bp} We follow here the same procedure as in Appendix~A of \ocite{prb2}; for example, for component $\rho_z$ of the density matrix we have
\be\begin{split}
& \frac{d\rho_z}{dt} = -i \langle\!\langle \left[\hat\sigma^z ; \delta\hat H \right] \rangle\!\rangle
= 2\tilde t s \langle\!\langle \sum_{a,b,\sigma} \left(\hat\sigma^+ - \hat\sigma^-\right) \\ & \left(\hat\tau^+ + \hat\tau^- \right)
\left(u^L_a u^R_b + v^L_a v^R_b \right)\left(\alpha^{L\dagger}_{a\sigma}\alpha^R_{b\sigma} - \alpha^{R\dagger}_{b\sigma}\alpha^L_{a\sigma}\right) \rangle\!\rangle
\end{split}\ee
All the quantities appearing in this equation are defined in Sec.~\ref{sec:model}.

The quantum statistical averages involving products of qubit and quasiparticle operators can be evaluated by solving their equation of motion in the Born approximation. In this way we find for instance
\begin{widetext}
\be\begin{split}
& \langle\!\langle \hat\sigma^+ \hat\tau^+ \hat\alpha^{L\dagger}_{a\sigma} \hat\alpha^R_{b\sigma}\rangle\!\rangle = i \tilde{t} \int_0^t \! d\tau \,
e^{i\left[\EZO - (\tilde\varepsilon_1 - \tilde\varepsilon_0)/2 +\epsilon_a^L - \epsilon_b^R +i0^+ \right](t-\tau)} \bigg\{
c_1 \left(u^L_a u^R_b -v^L_a v^R_b \right)\left(1-f^L_a\right) f^R_b \rho^o_+(\tau)
-c_0 \left(u^L_a u^R_b -v^L_a v^R_b \right) \\ & \times f^L_a \left(1-f^R_b\right) \rho^e_+(\tau)
-\frac{i}{4} s \left(u^L_a u^R_b +v^L_a v^R_b \right) \left[\left(1-f^L_a\right) f^R_b \left(1-\rho_z(\tau) -2\rho_0^z(\tau)\right) -
f^L_a \left(1-f^R_b\right)\left(1-\rho_z(\tau) +2\rho_1^z(\tau)\right) \right]
\bigg\} \, ,
\end{split}\ee
\end{widetext}
where we use the shorthand notation $f^j_a = f^j(\xi^j_a)$. Similar formulas can be obtained for all the density matrix components and all the quantum statistical averages determining their time evolutions. The procedure is lengthy but straightforward and leads, after introducing the Markov and rotating wave approximations as described in \ocite{prb2}, to the equations presented in Sec.~\ref{sec:me}.


\begin{thebibliography}{30}
\bibitem{noteswt}This reduction can involve suitable transformations and approximations that can take into account the role of virtual transition to higher levels.\cite{transmon}
\bibitem{footbw}This expectation is justified if the spectral density $S(\omega)$ varies slowly with frequency near $\EZO$; more precisey, if we indicate with $\Delta \omega$ the (smallest) distance in frequency from $\EZO$ over which $S(\omega)$ varies appreciably [$S(\EZO\pm\Delta\omega)/S(\EZO)\sim O(1)$], then the sufficient condition for the neglected terms to be small is $\varepsilon_1/\Delta\omega \ll 1$. Note that for typical $1/f$-like noise, with $S(\omega) \sim 1/\omega^\alpha$ and $\alpha \approx 1$ to 2, we have $\Delta \omega \simeq \EZO$; for this case the sufficient condition $\varepsilon_1/\EZO \ll 1$ is satisfied.
\bibitem{noteexc}This is also true for the excitation rates so long as $\delta E \gg |\varepsilon_1|$, which is the experimentally relevant regime as explained later in the text.
\bibitem{freqshift}As discussed in Refs.~\onlinecite{prl,prb}, quasiparticles cause a shift $\delta \omega$ in the qubit frequency. At leading order this frequency shift is independent of parity, as we have shown for the quasiparticle rates. Thus, we can assume that the frequency shift is accounted for by a small renormalization of $\EZO$ and do not write $\delta\omega$ explicitly in \esref{rpee}-\rref{rpoe} and those that follow.
\end{thebibliography}
\end{document}